\documentclass[9pt,twocolumn,twoside]{osajnl}

\journal{ol} 

\setboolean{shortarticle}{true}

\title{Chaotic time-delay signature suppression in lasers using phase-controlled dual optical feedback}

\author[1,*]{Robbe de Mey}
\author[1,2]{Spencer W. Jolly}
\author[3,4]{Alexandre Locquet}
\author[1]{Martin Virte}

\affil[1]{Brussels Photonics (B-PHOT), Dept. of Appl. Physics and Photonics (TONA), Vrije Universiteit Brussel (VUB), Pleinlaan 2, 1050 Brussels, Belgium}
\affil[2]{Service OPERA-Photonique, Université Libre de Bruxelles (ULB), Brussels, Belgium }
\affil[3]{IRL 2958 Georgia Tech-CNRS, Georgia Tech Lorraine, 2 Rue Marconi, 57070 Metz, France}
\affil[4]{School of Electrical and Computer Engineering, Georgia Institute of Technology, Atlanta, Georgia, 30332-0250, USA}

\affil[*]{Corresponding author: robbe.de.mey@vub.be}

\begin{abstract}
We experimentally study a semiconductor laser subject to two optical feedbacks in a free space setup. We show that the time delay signature, manifesting itself in the chaotic output intensity, can be better suppressed than in a laser with single feedback. We demonstrate that the control of the feedback phase is essential to suppress the time delay signature, in contrast to the one-delay case, and that the feedback phase also impacts the chaotic bandwidth. By optimizing the feedback phase the time delay signature can be reduced by a factor of 2 while maintaining a large chaotic bandwidth.
\end{abstract}

\setboolean{displaycopyright}{false}

\begin{document}
\maketitle

Optical feedback in semiconductor lasers can give rise to rich nonlinear behavior in terms of intensity, phase, and carrier dynamics \cite{Ohtsubo2013, Uchida2012, Donati2013, Tkach1985}. For example, feedback from a distant mirror can destabilize the laser leading to periodic, quasi-periodic, or chaotic behavior \cite{Ohtsubo2013, Uchida2012}. This chaotic behavior has been suggested for several practical applications, including, random number generation (RNG) \cite{Uchida2012, Reidler2009}, chaotic lidar \cite{Chembo2019}, and secure communication \cite{Uchida2012, Udaltsov2005}. For these applications, the pseudo-random chaotic dynamics generated by the semiconductor with optical feedback is crucial. However, the delay between laser and mirror, essential to trigger chaotic dynamics, also leaves a trace in the laser output. It is known as the Time Delay Signature (TDS); it reveals information about the system and is not desirable in applications. Indeed, for secure communications, the time delay is a crucial parameter that an eavesdropper could exploit \cite{Chembo2019,Udaltsov2003,Udaltsov2005,Ortin2005}; for chaotic lidar, the chaotic state should have a flat and smooth spectrum \cite{Lin2004}; for RNG, a strong delay signature is synonymous with failing randomness tests \cite{Ohtsubo2013,Uchida2012, Tang2015}. Suppressing the TDS in laser systems with optical feedback has been demonstrated in several ways: adjusting the injection current \cite{Rontani2007}, adjusting the feedback strength \cite{Wu2010}, using ring lasers \cite{Nguimdo2012}, using distributed feedback \cite{Song-SuiLi2015,Xu2017}, using polarization dependent feedback \cite{Lin2014} or adding quantum noise to the system \cite{Guo2021} to name a few. One straightforward way to reduce the TDS is achieved by adding a second optical feedback close to the first \cite{Wu2009,Lee2005}. Overall, adding one delay can either stabilize \cite{Tobbens2008,Liu1997,Ruiz-Oliveras2006,Rogister1999,Rogister2000} or further destabilize \cite{Tavakoli2020,Sukow2002,Onea2019,Barbosa2019} the laser depending on the specific configuration. In this case, the feedback system resembles an interferometer, known to be sensitive to phase changes. Indeed, earlier research indicates a feedback phase sensitivity of the dynamics \cite{Tobbens2008,Rogister1999,Rogister2000}. However, a small change in the mirror position, similar to a feedback phase change, was not yet investigated with respect to its impact on the TDS \cite{Wu2009}.
In this letter, we show that the feedback phase should be included as a control parameter as it plays a crucial role in the dynamics. Our experiments show that the relative feedback phase has a strong influence on the TDS suppression. First, we measure the threshold reduction due to one or two delays when changing the relative feedback phase. This gives an indication that the feedback phase plays a more important role for two delay systems. Second, we study the relation between the relative feedback phase and the TDS suppression and compare it again with the one-delay case.\\

\begin{figure}[b!]
\centering
\includegraphics[width=\linewidth]{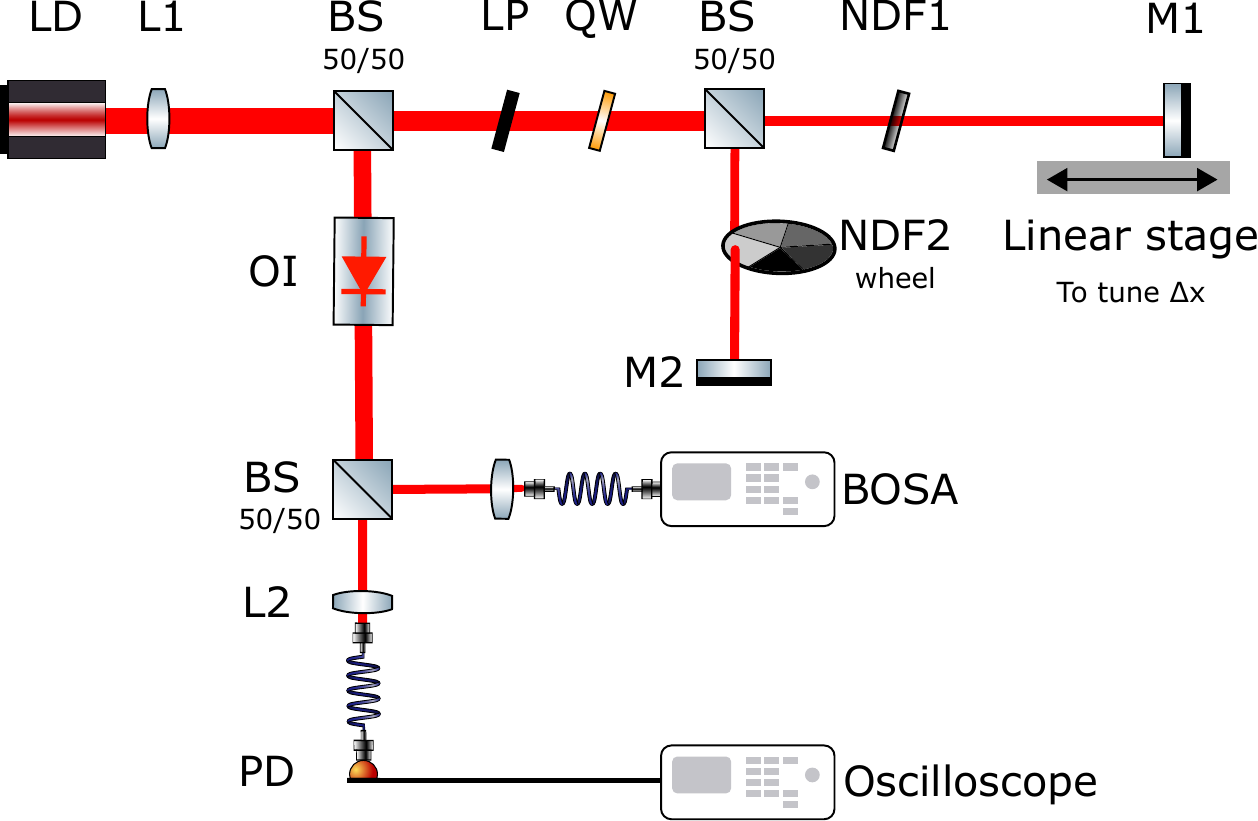}
\caption{Experimental setup. LD: Laser Diode, L: Lens, BS: Beam Splitter, LP: Linear Polarizer, QW: Quarter-Wave plate (rotatable), M: Mirror, NDF: Neutral Density Filter, OI: Optical Isolator, PD: photodetector.}
\label{fig:schematic}
\end{figure}

\begin{figure}[t]
	\centering
	\includegraphics[width=0.66\linewidth]{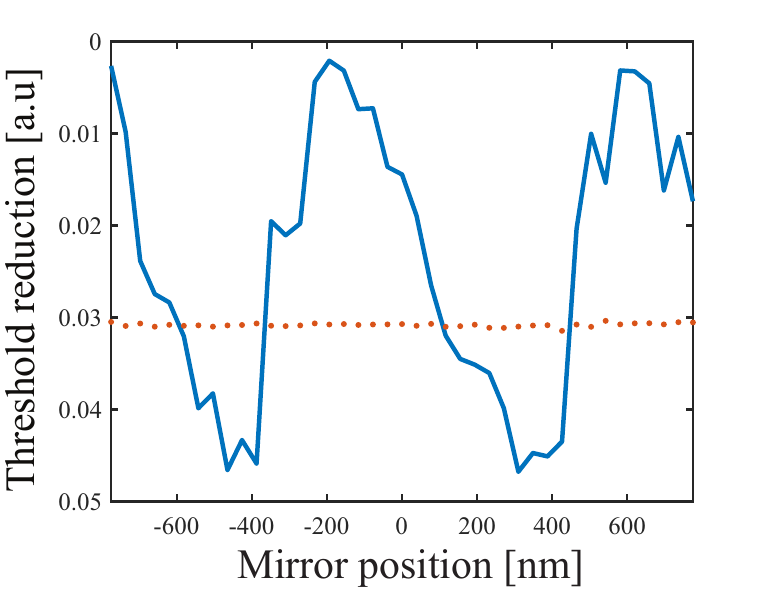}
	\caption{Relative threshold reduction versus mirror position for one delay (dotted orange) and two delays (continuous blue). }
	\label{fig:threshold}
\end{figure}

The experimental setup used is shown in figure \ref{fig:schematic}. A semiconductor laser is coupled to two mirrors. The laser is a commercially available single-mode edge emitting Distributed FeedBack (DFB) laser (3spTechnologies, 1953LCV1). The laser is kept at a constant temperature of $25^{\circ} C$. The mirror M1 in feedback loop 1 is on a linear stage (Newport XMS50) which can move in a range of 50 mm with a precision of 1 nm. By changing the position of the linear stage, we can therefore control the position of the mirror at the sub-wavelength scale and thus tune the feedback phase. The other mirror M2 remains fixed. We control the feedback strength in each feedback arm with Neutral Density Filters (NDFs) and the combination of a quarter-wave plate and a linear polarizer. As the light passes twice through the quarter-wave plate, it acts as a half-wave plate, rotating the polarization direction of the input linearly polarized light. By rotating the quarter-wave plate, in combination with the fixed polarizer, we obtain a variable optical attenuator. In this way, we fully control the feedback strength of both arms independently. An optical isolator separates the measurement arm from the rest of the system. The optical spectrum is measured with an Optical Spectrum Analyser with a resolution of 0.08 pm (Aragón Photonics BOSA). 
\begin{figure}[b!]
	\centering
	\includegraphics[width=\linewidth]{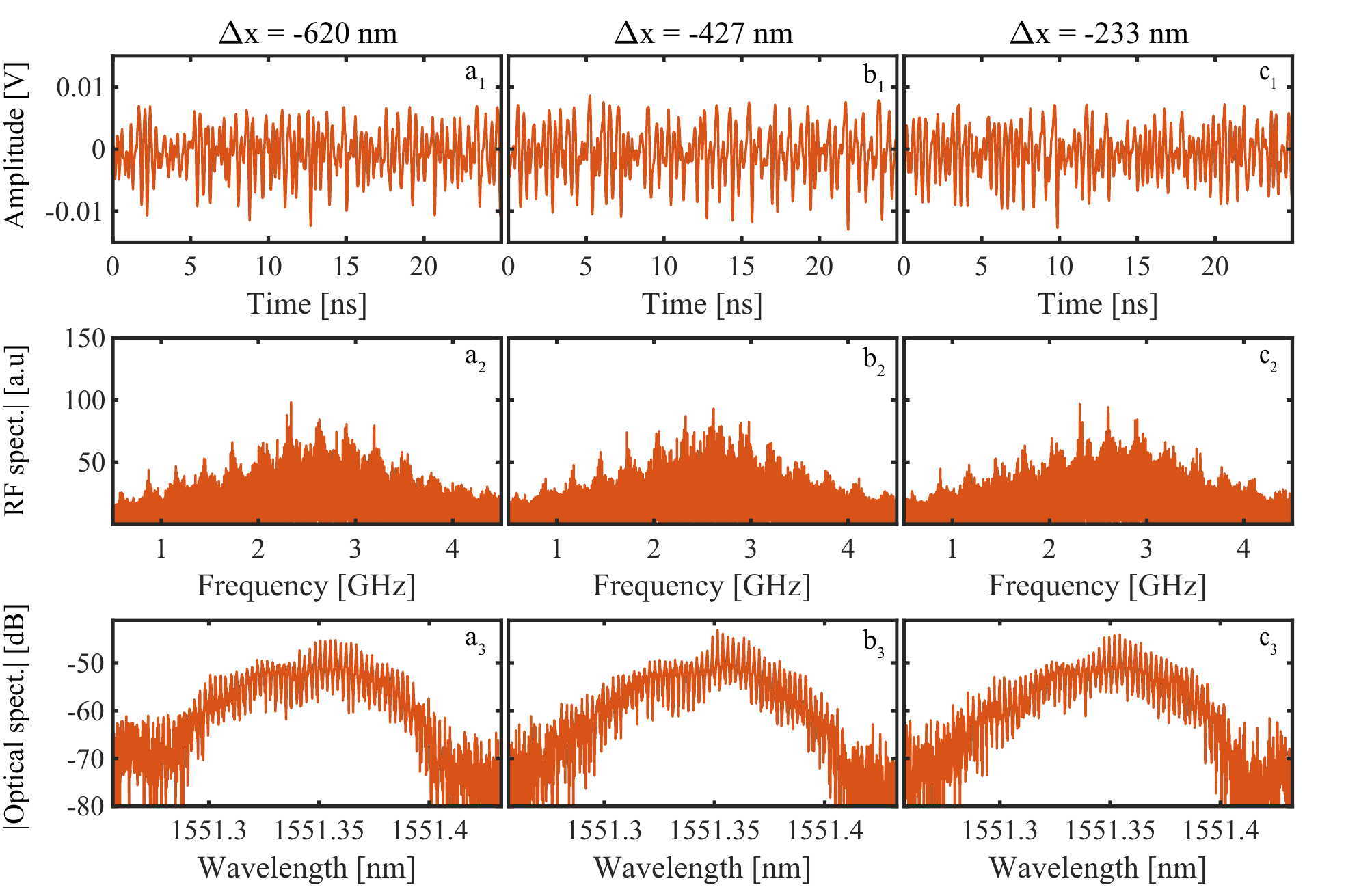}
	\caption{Laser coupled to a single mirror. From top to bottom: time series, RF spectrum (FFT), and optical spectrum. Each column is for a different position of mirror 1: a) -620 nm, b) -427 nm and c) -233 nm.  Changing the mirror position has a limited effect on the laser dynamics.}
	\label{fig:OneDelayTime}
\end{figure}
A 12 GHz photodiode (12 GHz New Focus 1544-B) is used to measure the time series of the output light with an oscilloscope (12 GHz bandwidth, Agilent DSO81204B). We use the DC bias voltage of the photodiode to measure the overall power of the light. Without feedback, the threshold current is $I_{th} = 21.3$ mA, and the laser emits single mode at 1551.34 nm. Unless stated otherwise, we keep the laser current at 30 mA, so $I_{th} = 1.4$. At this current, the relaxation oscillation frequency of the laser without feedback is $f_{RO} = 2.25$ GHz or $\tau_{RO} = 1/f_{RO} = 0.44$ ns.

\begin{figure}[b!]
	\centering
	\includegraphics[width=\linewidth]{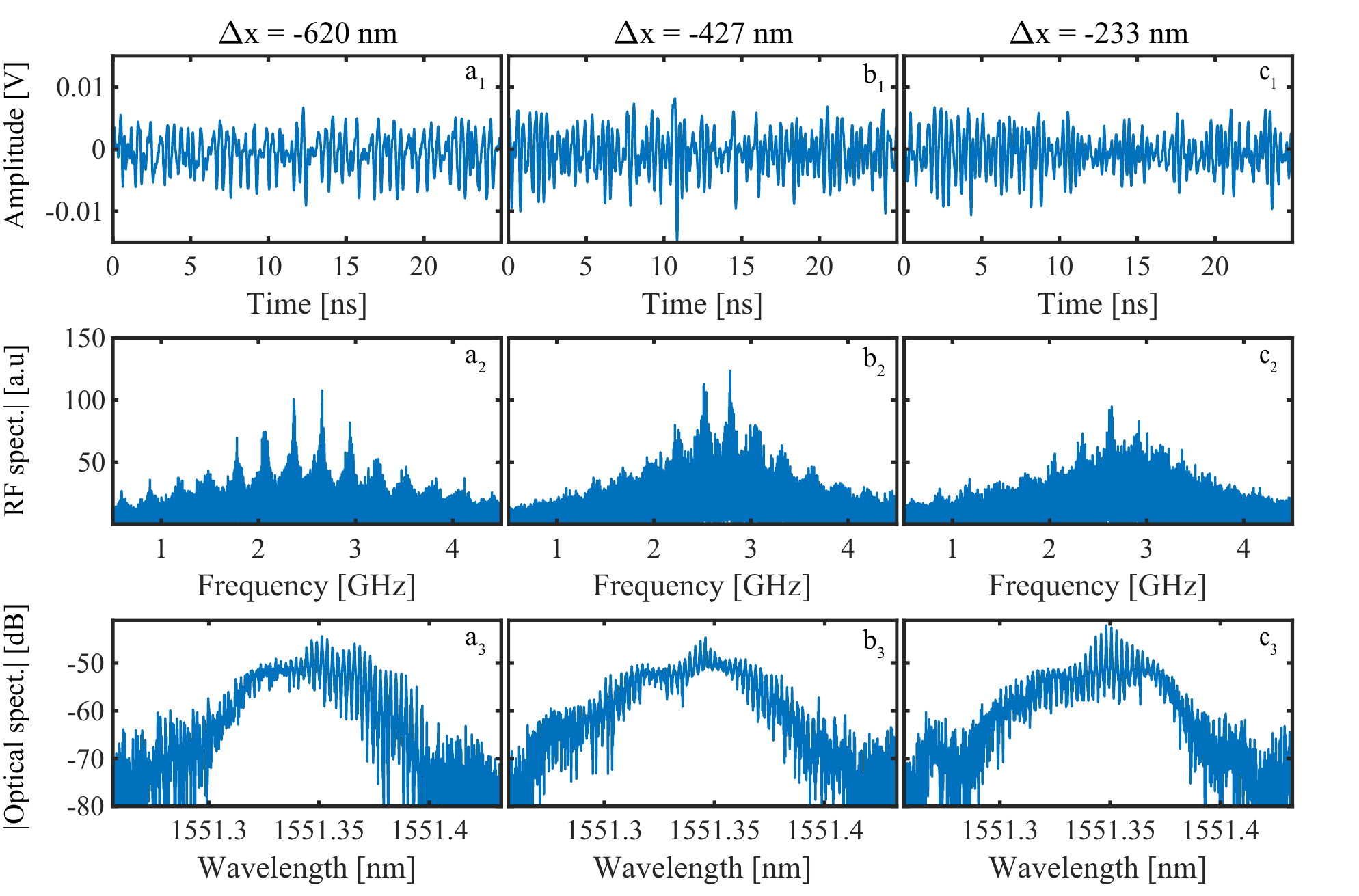}
	\caption{Laser coupled to two mirrors. From top to bottom: time series, RF spectrum (FFT), and optical spectrum. Each column is for a different position of mirror 1: a) -620 nm, b) -427 nm and c) -233 nm.  Changing the mirror position has a stronger effect on the laser dynamics than in the case of a single feedback.}
	\label{fig:TwoDelayTime}
\end{figure}

Both mirrors are set approximately at the same distance from the laser cavity. We estimate the delay by measuring the distance between peaks in the RF spectrum of the intensity: for feedback loop 1, this gives $\tau_1 = 3.35$ ns, and for feedback loop 2, $\tau_2  = 3.42$ ns, thus a difference of 0.07 ns, i.e. $0.16 \tau_{RO}$.  In ref. \cite{Wu2009} it is proposed that for a difference of $0.5 \tau_{RO}$ a minimum of the TDS peak is reached. Here the difference is slightly smaller, but we will still observe stronger suppression than with only one delay. We then change the mirror position in small increments around $\tau_1$. The total displacement of mirror 1 is 1551 nm, decomposed in 41 identical steps. Therefore, over this range, we tune the relative feedback phase between the feedback arms by $4\pi$ (two full periods). We measure the threshold reduction for each arm (blocking the other) to have an indication of the feedback strength through the relative threshold current: $\Delta i = (I_{th} - I)/I_{th}$ with $I_{th}$, and $I$ the threshold currents without and with feedback, respectively \cite{Hohl1999}. For the experiment with two delays, the relative threshold reductions are $\Delta i_1 = 0.017$ and $\Delta i_2 = 0.015$, indicating that the feedback strength of loop 2 is slightly lower than in loop 1. 
We also measure the relative threshold reduction with both feedback arms open for each position of mirror M1 and obtain the results shown in figure \ref{fig:threshold}. There is a clear periodic dependence between the relative threshold reduction and the feedback phase, which suggests that interference is occurring in the feedback loops. Depending on the relative feedback phase, the feedback light can either interfere constructively or destructively. This is a qualitative difference with the one-delay case as such effect cannot occur with a single feedback loop. Indeed, we performed the same experiment for the one-delay case. To isolate the effect of the feedback phase, we adjust the feedback strength so that the same amount of light is fed back inside the laser cavity as in the two-delay case. The determination is based on the threshold reduction. Specifically, by adjusting the NDFs, we set the relative threshold reduction to 0.03, which is close to the sum of the threshold $\Delta i_1$  and $\Delta i_2$.  As shown in figure \ref{fig:threshold}, no clear dependence with respect to the feedback phase is visible. \\

We now turn to the main experiment to investigate the impact of the feedback phase on the TDS. The injection current is 30 mA, all other parameters are the same as in the threshold reduction measurement. Figure \ref{fig:OneDelayTime} shows the results of the experiment with one optical feedback for three different positions of mirror 1, -620 nm, -427 nm, and -233 nm (or in terms of feedback phase: -5.0, -3.5, and -1.9 radians). The time series, RF spectrum (calculated by taking the FFT of the time series), and optical spectrum are shown. Figure \ref{fig:TwoDelayTime} shows the same experiment but for the case of double optical feedback. From the time series (figure \ref{fig:OneDelayTime} (a\textsubscript{1}, b\textsubscript{1}, and c\textsubscript{1}) and figure \ref{fig:TwoDelayTime} (a\textsubscript{1}, b\textsubscript{1}, and c\textsubscript{1}) both cases are chaotic for all mirror positions. Moreover, the two cases seem similar based on their time series. Yet, from the RF spectrum, it can be seen that the two-delay case is more sensitive to the feedback phase, see figure \ref{fig:OneDelayTime} (a\textsubscript{2}, b\textsubscript{2}, and c\textsubscript{2}) and figure \ref{fig:TwoDelayTime} (a\textsubscript{2}, b\textsubscript{2}, and c\textsubscript{2}). Overall, the RF spectrum is broad and displays peaks separated by approximately $1/\tau_1$. For the one-delay case, changing the mirror position seems to have only a minor impact on the RF spectrum. On the other hand, for the two-delay case, the RF spectrum changes more significantly with the mirror position. In figure \ref{fig:TwoDelayTime} (a\textsubscript{2}) clear peaks appear, which are suppressed when changing the mirror position to -233 nm as in figure \ref{fig:TwoDelayTime} c\textsubscript{2}. When looking at the optical spectrum in figure \ref{fig:OneDelayTime} (a\textsubscript{3}, b\textsubscript{3}, and c\textsubscript{3}) and figure \ref{fig:TwoDelayTime} (a\textsubscript{3}, b\textsubscript{3}, and c\textsubscript{3}), a similar trend occurs though in a more subtle manner than in the RF spectra. The optical spectrum does not change much for the one-delay case, while more significant changes can be observed in the two-delay case, in particular the width of the spectrum.

\begin{figure}
\centering
\includegraphics[width=\linewidth]{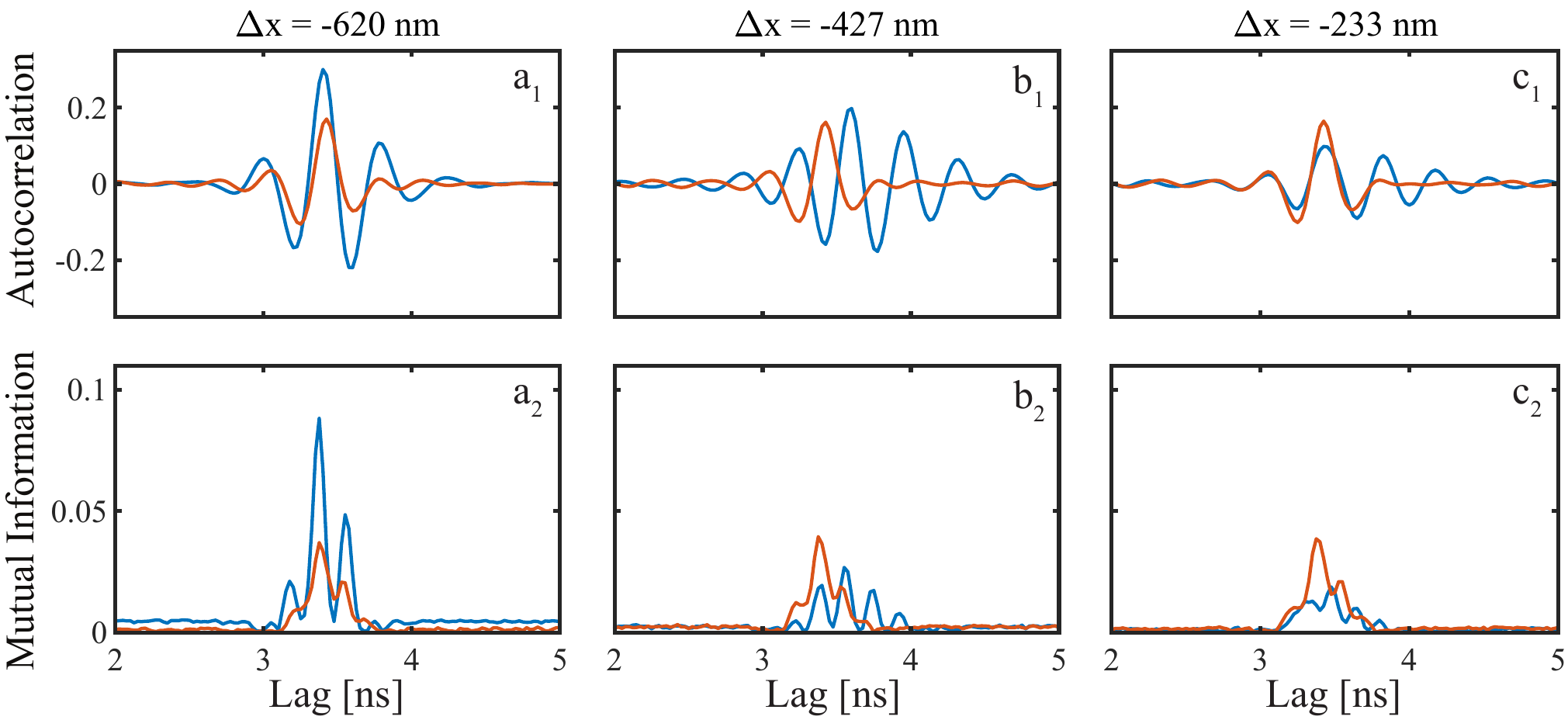}
\caption{Comparison of the TDS between the one- (orange) and two-delay cases (blue) for three different mirror position. Top: autocorrelation function. Bottom: delayed mutual information.}
\label{fig:oneAndTwoDelayAutocorr}
\end{figure}

These results show that the two-delay system dynamics depend more strongly on the feedback phase. To study how this influences the TDS we calculate the autocorrelation function (ACF) $\rho$ and Delayed Mutual Information (DMI) from the time series of the intensity of the laser \cite{Rontani2009}. Figure \ref{fig:oneAndTwoDelayAutocorr} shows the ACF and DMI of the one- and two-delay cases for the same three mirror positions as before. For both figures of merit, the height of the peak quantifies the prominence of the TDS. For the one-delay case, only minor changes are visible when the feedback phase is tuned. But for the two-delay case the TDS is clearly impacted. Changing the feedback phase changes the shape, height, and position of the TDS peak. Moreover, for a mirror displacement $\Delta x = -620$ nm (figure \ref{fig:oneAndTwoDelayAutocorr} (a\textsubscript{1} and a\textsubscript{2})), the dual delay case has a more pronounced TDS, while for $\Delta x = -233$ nm (figure \ref{fig:oneAndTwoDelayAutocorr} (c\textsubscript{1} and c\textsubscript{2})), the TDS is more suppressed than for the one-delay case. Moreover, for the two-delay case, both functions display a rather flat bump around the delay value rather than a sharp peak, making it more difficult to identify accurately the time-delay of the system. The results clearly show that for a semiconductor laser with only one optical feedback, changing the mirror position on this scale does not significantly affect the dynamics. The RF spectrum, optical spectrum, ACF, and DMI are almost unchanged. All observed variations can likely be attributed to noise.\\

\begin{figure}
\centering
\includegraphics[width=\linewidth]{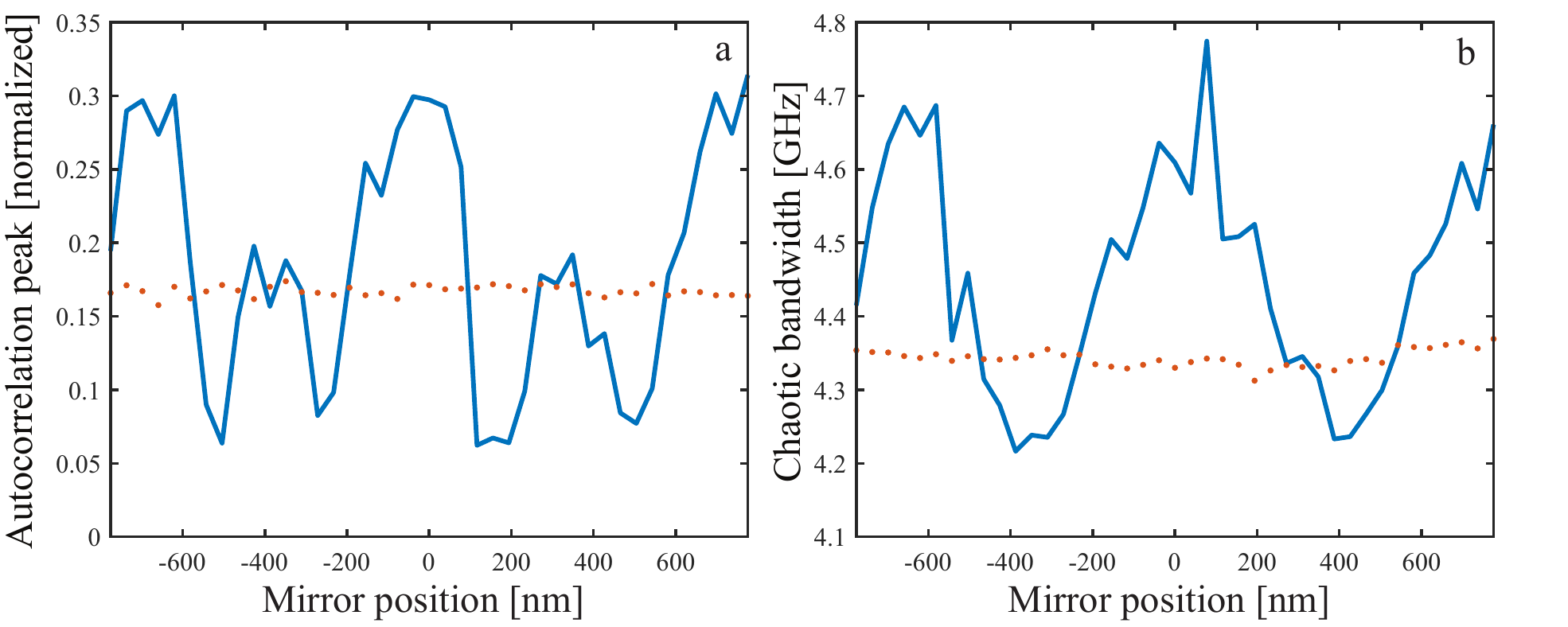}
\caption{Comparison of (a) height of the largest peak in the autocorrelation function and (b) chaotic bandwidth for double optical feedback (blue) and the equivalent one-delay case (orange).}
\label{fig:autoCorr}
\end{figure}

The above results demonstrate that the suppression of the TDS with a second delay at a slightly different time-delay and strength from the first is only feasible when the feedback phase can be controlled. To investigate the effect systematically, we look at the height of the largest peak of the absolute ACF for all the feedback phase steps in the experiment, shown in figure \ref{fig:autoCorr} (a). For each mirror position, and thus each feedback phase, we calculate the height of the peak in the absolute ACF in a range around the lag where we expect the TDS, specifically in $3.3 \pm 1.0$ ns. For the two-delay case, the height of the TDS changes significantly. We find a minimum at $\Delta x = 116$ nm with $|\rho| = 0.06$. The maximum is $|\rho| = 0.31$ at $\Delta x = 775 $nm.  For the one-delay case, the peak stays around $|\rho| = 0.18$. Similar trends are observed in the DMI. Therefore, by sweeping the feedback phase in the two-delay case, the TDS can either be further suppressed or increased. 
For the sake of completeness, we performed a systematic search in the experimentally achievable parameter space for the case of only one feedback to find the optimal value for the TDS. By changing the current, delay, and feedback strength, the maximum reduction of the TDS corresponds to $|\rho| = 0.15$ in the one delay case. It thus appears that adding a second delay can further suppress the TDS by a factor of 2, but, again, this is only possible with feedback phase control. \\

Our results shed new light on the observations of ref. \cite{Wu2009}. We confirm that a second optical feedback can indeed suppress the TDS more than what  can be achieved with only one feedback. However, our results also show that the relative feedback phase, associated to small sub-wavelength variations of the mirror positions, has a strong influence on the suppression. For a change of only 200 nm in mirror position, the height of the TDS peak can change significantly, from optimal to worst suppression. We can therefore conclude that, to use the two-mirror approach to effectively suppress the TDS, the relative feedback phase should be taken into account to avoid significant degradation of the concealment of the TDS. 
Besides the effect on the TDS, the RF spectrum also changes when tuning the feedback phase in the two-delay case, as seen in figure \ref{fig:TwoDelayTime} (a\textsubscript{2}, b\textsubscript{2}, c\textsubscript{2}). To quantify this, we analyse the chaotic bandwidth (CHBW), defined in \cite{Lin2012}. Figure \ref{fig:autoCorr} (b) shows how the CHBW changes when changing the feedback phase for both the one- and two-delay cases. Again, for the one-delay case, the CHBW remains almost constant with a value just below 4.4 GHz, while for the two-delay case, there is a clear dependence on the relative feedback phase. The mean value for the two-delay case is 4.45 GHz, slightly higher than the equivalent one-delay value. The maximum value is 4.77 GHz and occurs for $\Delta x = 77$ nm.
The comparison of figure \ref{fig:autoCorr}(a) and (b) shows that the maximum TDS suppression and maximum CHBW do not occur for the same mirror position. For the two-feedback system, there seems to be a trade-off between TDS and CHBW, and one cannot optimize them together. However, it is possible to have both a stronger suppression and a higher chaotic bandwidth compared to the one-delay case, for example, around mirror position 200 nm in figure \ref{fig:autoCorr}. In addition, the gain in TDS suppression is more significant than the loss in chaotic bandwidth. Specifically, the minimum TDS is 2.5 times smaller than the mean, but the minimum bandwidth is only a few percent lower than the mean. So, although they cannot be fully optimized together, the bandwidth reduction remains minor.\\

To conclude, we show that the relative feedback phase or small subwavelength changes of the mirrors’ position are important to understand the dynamics of a semiconductor laser coupled to two optical feedback loops. We demonstrate that the threshold reduction, a measure for the feedback strength, is strongly influenced by the feedback phase. Moreover, the increased sensitivity to a small change in the mirror position manifests itself strongly in the chaotic dynamics. We confirm that a second delay close to the first, on a scale corresponding to the relaxation oscillation period \cite{Wu2009}, can further suppress the TDS compared to the equivalent one-delay case. However, our experimental results also show that it is crucial to control the position of the mirrors very precisely. Sweeping the delay on the scale of  the optical wavelength has a strong influence on the feedback strength, the time delay signature and the chaotic bandwidth. Our experiments show that we can further suppress the TDS compared to the best suppression in the one-delay case.\\

\begin{backmatter}
\bmsection{Funding}Fonds Wetenschappelijk Onderzoek (FWO), project G0E7719N. Vlaamse Overheid, METHUSALEM program. FWO and embassy of France in Belgium, TOURNESOL program. Conseil Régional Grand Est. European Union H2020 research and innovation program, Marie Sklodowska-Curie Action (MSCA), project 801505. 

\bmsection{Disclosures} The authors declare no conflicts of interest.

\bmsection{Data Availability Statement} Data underlying the results presented in this paper are not publicly available at this time but may be obtained from the authors upon reasonable request.

\end{backmatter}

\bibliography{sample}

\end{document}